\documentclass[twocolumn,showpacs,preprintnumbers,amsmath,amssymb,bibnotes]{revtex4}

\usepackage{graphicx}
\usepackage{dcolumn}
\usepackage{bm}

\begin{document}


\title{Squeezed state purification with linear optics and feed forward}

\author{O. Gl\"ockl$^1$}
\email{gloeckl@kerr.physik.uni-erlangen.de}
\altaffiliation[present address: ]{ARC Centre of Excellence for Quantum Atom Optics, The Australian National University, Canberra ACT 0200, Australia}
\author{U. L. Andersen$^1$}
\author{R. Filip$^{1,2}$}
\author{W. P. Bowen$^3$}
\author{G. Leuchs$^1$}
\affiliation{%
$^1$ Institut f\"ur Optik, Information und Photonik, Max--Planck
Forschungsgruppe, Universit\"at Erlangen--N\"urnberg,\\
G\"unther--Scharowsky--Stra{\ss}e 1 / Bau 24, 91058 Erlangen, Germany\\
$^2$ Department of Optics, Palacky University, Olomouc, Czech Republic,\\
$^3$ Department of Physics, University of Otago, Dunedin, New Zealand}%

\date{\today}

\begin{abstract}

A scheme for optimal and deterministic linear optical purification of mixed squeezed Gaussian states is proposed and
experimentally demonstrated. The scheme requires only linear
optical elements and homodyne detectors, and allows the balance
between purification efficacy and squeezing degradation to be
controlled. One particular choice of parameters gave a ten-fold reduction of the thermal noise
with a corresponding squeezing degradation of only
11~\%. We prove optimality of the protocol, and show that it can be used to enhance the performance of quantum informational protocols such as dense coding and
entanglement generation.
\end{abstract}

\maketitle


Quantum information science facilitates many information processing tasks that were previously thought intractable or impossible \cite{nielsen}. Without exception, however, these protocols are highly susceptible to decoherence and impurities \cite{computing}. As a result, significant effort has been put into developing procedures to purify the required non-classical resources.
It has recently been shown that perfect single photon purification is impossible using only linear optics, rendering the protocol rather challenging \cite{berry04.pra}. Nevertheless, techniques to non-deterministically purify single photon states (as a resource) have been
proposed both when a single copy and two copies are available \cite{munro-konrad}, but so far, no experiments have been reported. Here, we examine the complementary problem of purification of continuous variable (CV) states, in particular Gaussian squeezed states. We find that in this regime, in contrast to the single photon case, purification of CV resource states is indeed achieveable with linear optics.

Gaussian squeezed states are an essential resource for deterministic quantum information processing; allowing the immediate generation of CV entanglement \cite{ent}, and
facilitating a variety of important protocols such as
unconditional quantum teleportation \cite{tel}, dense coding
\cite{braunstein-li}, and entanglement swapping \cite{LOO99}.  The
efficacy of these protocols typically depends critically on both
the degree of squeezing and the purity of the squeezed states.
Presently available squeezed states are often highly impure (i.e.
mixed) due to decoherence and dissipation in the preparation
sources.  It is therefore important to develop effective
purification techniques for these states.  In this paper, we
determine the optimal level of purification possible for single
Gaussian squeezed states, propose an experiment that  achieves
this level, and perform an experimental demonstration.  In
contrast to single photon purification, the protocol is
deterministic, which presents significant advantages in terms of
scalability and reliability in compound quantum information
protocols.  Additionally, although the past decade has seen
significant improvements in the degree of squeezing achievable by
experimentalists \cite{BACH}, the work present here represents the
first experiment dedicated to improving purity.

The purity of a Gaussian squeezed state is given simply by
$tr(\rho ^2)=1/\sqrt{\Delta ^{\! 2} \! \hat X\cdot\Delta^{\! 2} \!
\hat Y}$ \cite{Paris}, where $\Delta^{\! 2} \! \hat X$ and $\Delta^{\! 2} \!
\hat Y$ are the variances of  the squeezed [$\hat X = \hat a^\dagger +\hat a$] and  the anti-squeezed [$\hat Y=i(\hat a^\dagger-\hat a)$] quadratures,
respectively, and $\rho$ is the density matrix.  A maximally pure state has $tr(\rho ^2)=1$, however, in realistic systems the purity is typically much worse. In fibre
based systems, for example, guided-acoustic-wave Brillouin
scattering causes $tr(\rho ^2)$ to be typically smaller than $0.1$
\cite{SHE86}.  Experimentally feasible squeezed state purification
techniques that do not require non-linear interactions or non-classical 
ancilla states are vital for these states to be useful in quantum information systems. 

In the Heisenberg picture a generic Gaussian operation can be described by the linear transformations \cite{holevo} 
\begin{equation}
\hat X'=\nu \hat X + \hat X_N, \quad \hat Y'=\mu \hat Y + \hat Y_N,
\label{trans}
\end{equation}
where $\nu ,\mu \in \Re$, and $\hat X_N$ and $\hat Y_N$ are
operators associated with noise added by the device in the
amplitude and phase quadratures, respectively. We see from the
form of $tr(\rho ^2)$ that the optimal purification technique must
minimize $\Delta^{\! 2} \! \hat Y$ whilst minimally disturbing the
degree of squeezing $\Delta^{\! 2} \! \hat X$.  Let us therefore
identify the minimum achievable $\Delta^{\! 2} \! \hat Y'$ for a
given $\Delta^{\! 2} \! \hat X'$. Assuming no correlations between
the input and noise operators, the amplitude and phase quadrature
variances of the purified state are given by
\begin{equation}\label{eqn-variance1}
\Delta^{\! 2} \! \hat X'=\nu^2 \Delta^{\! 2} \! \hat X+\Delta^{\!
2} \! \hat X_\mathrm{N}, \! \! \quad \Delta^{\! 2} \! \hat
Y'=\mu^2 \Delta^{\! 2} \! \hat Y+ \Delta^{\! 2} \! \hat
Y_\mathrm{N}.
\end{equation}
From (\ref{trans}) and the quadrature operator commutation relations ($[X,Y]=[X',Y']=2i$) we also find $[X_N,Y_N]=2i(1-\mu\nu)$ and thus the uncertainty relation 
$\Delta^{\! 2} \! \hat X_\mathrm{N} \Delta^{\! 2} \! \hat Y_\mathrm{N}\geq(1-\nu\mu)^2.$
A lower bound on $\Delta^{\! 2} \! \hat Y'$ can be obtained from
the saturation of this inequality and
Eqs.~(\ref{eqn-variance1})\footnote{It is assumed here that
$\Delta^{\! 2} \! \hat X_\mathrm{N} \neq 0$.}
\begin{equation}\label{eqn-minvar}
\Delta^{\! 2} \! \hat
Y_{\min}'(\nu,\mu)=\frac{(1-\nu\mu)^2}{\Delta^{\! 2} \! \hat
X'-\nu^2 \Delta^{\! 2} \! \hat X}+\mu^2 \Delta^{\! 2} \! \hat Y.
\end{equation}
Our goal is to determine the minimum of Eq.~(\ref{eqn-minvar})
with respect to $\nu$ and $\mu$, achievable using linear
transformations.  The range of $\nu$ can be constrained. From Eqs.~(\ref{eqn-variance1}) we have
$\nu^2 < {\Delta^{\! 2} \! \hat X'}/{\Delta^{\! 2} \! \hat X}$.
Furthermore, since only linear operations are allowed an input
coherent state cannot be output squeezed. As a
result $\Delta^{\! 2} \! \hat X_\mathrm{N}\geq1-\nu^2$, and
therefore $\nu^2 \ge (1-\Delta^{\! 2} \! \hat X')/(1-\Delta^{\! 2}
\! \hat X)$.
The minimum of Eq.~(\ref{eqn-minvar}) can be found by first minimizing over $\mu$. It is then easy to see that the minimum within our constraints on $\nu$ occurs at the lower bound with $\nu^2=(1-\Delta^{\! 2} \! \hat X')/(1-\Delta^{\! 2}\! \hat X)$, and we find
\begin{equation}\label{eqn-minvar2}
\Delta^{\! 2} \! \hat Y_{\min}'=\frac{\Delta^{\! 2} \! \hat Y
(1-\Delta^{\! 2} \! \hat X)}{(1-\Delta^{\! 2} \! \hat
X')+\Delta^{\! 2} \! \hat Y(\Delta^{\! 2} \! \hat X'- \Delta^{\!
2} \! \hat X)}.
\end{equation}
This expression determines the optimal solution for a deterministic linear purification scheme \footnote{Another potential solution for $\Delta^{\! 2} \! \hat Y'$ exists for $\Delta^{\! 2} \! \hat X_\mathrm{N} = 0$ (this corresponds to noiseless amplification \cite{BUC98}). However, the solution is worse than that found in Eq.~(\ref{eqn-minvar2}).}.

\begin{figure}
\includegraphics[scale=.5]{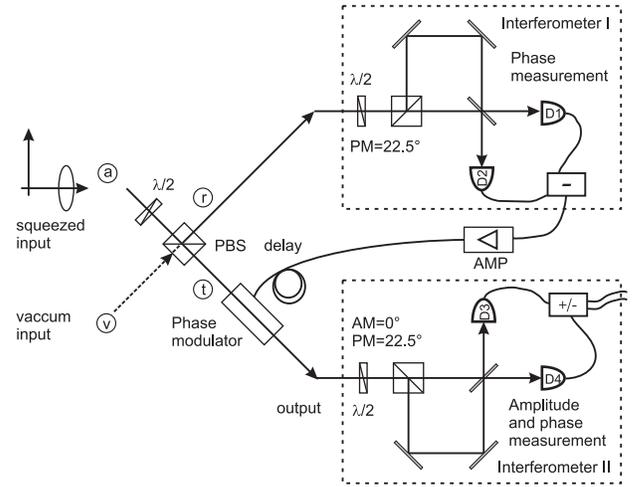}
\caption{\label{bild-aufbau} Experimental setup of the phase noise
subtraction scheme. A fraction of input mode $\hat a$ is tapped
off to perform phase quadrature measurements using interferometer
I. The resulting photocurrent is amplified (AMP) and, after
introducing an appropriate cable delay, fed forward to the
modulator. The amplitude and phase noise of the transmitted mode
is characterized by interferometer II. AMP: electrical amplifier,
PBS: Polarizing beam splitter, $\lambda/2$: Half wave plate.}
\end{figure}

Let us now consider the optimality of the purification scheme shown in Fig.~\ref{bild-aufbau}. The basic idea is to use homodyne detection and electro optic feed forward to control the quantum fluctuations similar to earlier work reported e.~g.\ in \cite{feedforward}. A fraction
$\sqrt{1-\eta}$ of the amplitude squeezed input mode $\hat a$ is
tapped off at a variable beam splitter consisting of a polarizing
beam splitter and a half-wave plate.
The (anti-squeezed) phase quadrature of the reflected beam $\hat
r$ is measured. The resulting measurement outcome is amplified by
a gain factor $g$, and fed forward onto the transmitted mode $\hat
t$ using a phase modulator. By optimizing the gain a minimum phase
quadrature variance of $\Delta^{\! 2} \! \hat
Y_{t,\min}=\Delta^{\! 2} \! \hat Y_\mathrm{a}/(\eta+(1-\eta)
\Delta^{\! 2} \! \hat Y_\mathrm{a})$ can be achieved for the
transmitted beam, where $\Delta^{\! 2} \! \hat Y_\mathrm{a}$
denotes the phase quadrature variance of the input.
Since the amplitude and phase quadratures are orthogonal, the
feed-forward will not affect the amplitude quadrature of the
transmitted mode, the variance of which is therefore simply
$\Delta^{\! 2} \! \hat X_\mathrm{t}=\eta \Delta^{\! 2} \! \hat
X_\mathrm{a} +1-\eta$. Solving this relation for $\eta$ and
substituting into the relation for $\Delta^{\! 2} \! \hat Y_{\rm
t,min}$ we find that this device delivers the desired optimum
performance described by (\ref{eqn-minvar2}).

In our experiment, we use intense amplitude squeezed beams from a
fibre Sagnac interferometer \cite{schmitt-kry-fio}. To perform
phase quadrature measurements we use an asymmetric Mach Zehnder
interferometer \cite{GLO04} which translates phase noise into
directly detectable amplitude noise at a specific sideband
frequency. Amplitude noise measurements are accomplished by
rotating the input polarization by 45$^\circ$, thus bypassing the
interferometer.

\begin{figure}\begin{center}
\includegraphics[scale=.95]{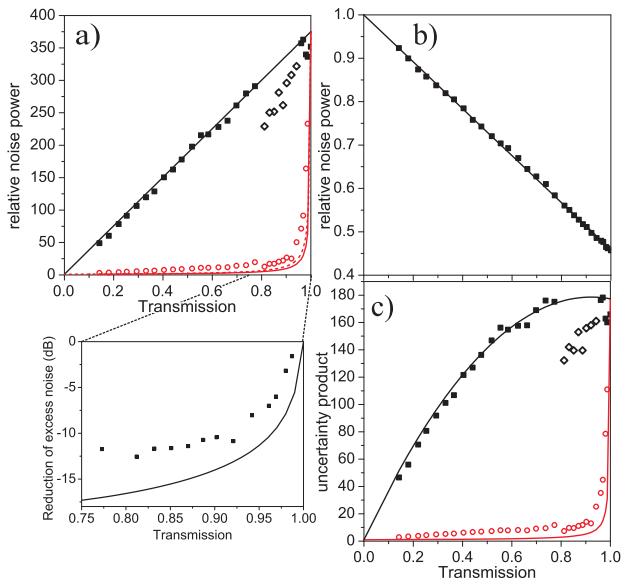}
\caption{Experimental results as a function of transmission. a)
and b): phase and amplitude quadrature variances of the transmitted beam,
c): product of phase and amplitude quadrature
variances.  Open circles and filled squares represent the results
with and without feed-forward, respectively; open diamonds in a) represent values taken at reduced detection efficiency (cf. main text). Also plotted is the degree of excess noise 
reduction which is independent of the actual detection efficiency. The lines are
obtained from our theoretical model, with the dashed line in a)
taking into account nonunit phase quadrature measurement efficiency of mode $\hat r$. All variances were normalized to the
shot noise level. Measurement frequency: $20.5$~MHz. Electronic
noise of $-83.3$~dBm, was subtracted from the measured data.
}
\label{bild-messpurification} \end{center}
\end{figure}

Since all involved states are Gaussian, the performance of the
device is fully characterized by measuring the variances
$\Delta^{\! 2} \! \hat X_\mathrm{t}$ and $\Delta^{\! 2} \! \hat
Y_\mathrm{t}$ of the transmitted state. These variances are
measured using interferometer II with and without the use of the
feed-forward loop. The results and theoretical predictions are plotted as a function of  the transmission $\eta$ in Fig.~\ref{bild-messpurification}. The overall detection efficiency $\eta_\mathrm{Det}$ of the Mach Zehnder interferometers used for phase measurement was around $70$\% due to imperfect photo diodes, interference contrast, and losses in optical components. Non-unity detection efficiency affects the detected noise levels. As the measured quantities in both cases (that is, with and without feed forward) are well above the shot noise limit the vacuum noise contribution due to the inefficiency can be neglected, and the measured variances therefore almost exactly scale with the efficiency of the verification stage. Hence, the ratio $\Delta^2\hat Y_{\mathrm t}/\Delta^2\hat Y_{\mathrm t,\min}$ yields a measure of the excess noise reduction that is to good approximation independent of the detection efficiency. When the introduction of vacuum noise is taken into account, we see that any degradation of the detection efficiency can only result in an underestimation of the level of purification. 

The experimentally observed phase quadrature variance of the transmitted beam is shown in Fig.~\ref{bild-messpurification}a. When the feed-forward was switched off (filled squares), the phase quadrature variance $\Delta^2 \hat Y_{\mathrm t}$ decreased linearly as the transmission was reduced simply due to the beam splitter attenuation. The solid line shows the expected behavior of the measured phase noise as a function of $\eta$. When the feed-forward was active (open circles) the behavior changed dramatically, with the phase quadrature variance $\Delta^2 \hat Y_{\mathrm t,\min}$ decreasing much more rapidly. The predicted behavior is plotted as a function of transmission (gray solid line). The detection efficiency of interferometer I limits the performance of the scheme (indicated by the dashed gray line). During the measurement run, the visibility, and hence the detection efficiency of interferometer II, gradually degraded. Finally, at a transmission $\eta=0.81$, interferometer II was realigned giving rise to a discontinuity in the measured noise levels. However, as can be seen from the enlargement of Fig.~\ref{bild-messpurification}a and as discussed above, the purification process is to good approximation independent of detection efficiency, and can certainly not be enhanced by worsening efficiency.

Fig.~\ref{bild-messpurification}b shows the amplitude quadrature
variance $\Delta^2 \hat X_{\mathrm t}$ as a function of transmission. The level of squeezing
decreases linearly as $\eta$ is reduced, as is typical of squeezed states upon attenuation.  As
expected, identical behavior was observed with and without feed-forward. From the interpolated variance at $\eta \rightarrow0$ and that measured at $\eta=1$, we conclude that $3.4$~dB of squeezing was present in the detected input field.

The uncertainty product (which is
directly related to the purity) of the transmitted state is plotted as a function of
$\eta$ in Fig.~\ref{bild-messpurification}c. The
open circles and filled squares correspond to the experimental
results with and without the use of feedforward, respectively. The
expected theoretical behavior is again plotted for both cases, and
agrees well with the experimental results. It is clear that the
purity can be increased considerably, even for quite high
transmission ($>90$~\%).  When $\eta = 0.92$, for example, the
purity is improved by more than an order of magnitude, while the
squeezing is only degraded by 11~\% from $\Delta^{\! 2} \! \hat
X_{\rm t} = 0.47$ to $\Delta^{\! 2} \! \hat X_{\rm t} = 0.52$.

\begin{figure}[t]
   \includegraphics[width=7.6cm]{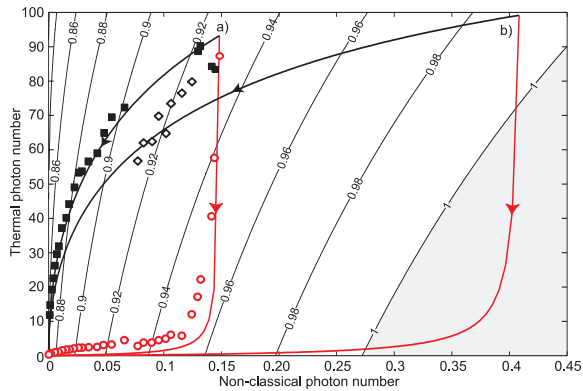}
   \caption{\label{PhotonNumberDiagram} Photon number diagram showing
experimental purification results and ideal purification
trajectories.  a) ideal trajectory for initial state used in
experiments.    b) trajectory for initial state with $\Delta^{\!
2} \! \hat X_{\rm sqz}=0.3$ and $\Delta^{\! 2} \! \hat Y_{\rm
anti}=400$. Dark trajectories/points: no feed-forward. Light
trajectories/points: feed-forward.  The contours represent lines
of constant $C_{\rm EPR}$   for $\langle \hat n \rangle_{\rm cp}
= 400$, normalized to $C_{\rm Holevo}$. Shading indicates the region
in which dense coding is successful.}
\end{figure}

Mechanisms that produce squeezed states of light rely on the
generation of photons in correlated pairs. This photon pairing is
responsible for the non-classicality of the states, while impurity
is caused by unpaired photons.  From this perspective, the effect
of our purification scheme is to selectively remove unpaired
(thermal) photons, while retaining the maximal number of pairs.
This process is visualized on the photon number diagram in
Fig.~\ref{PhotonNumberDiagram}, where the mean number of
non-classical $\langle \hat n \rangle_\mathrm{non-cl}$ and thermal
$\langle \hat n \rangle_\mathrm{thermal}$ photons per bandwidth
per time are calculated from $\langle \hat n
\rangle_\mathrm{non-cl}= (\Delta^{\! 2} \! \hat
X_\mathrm{sqz}+1/\Delta^{\! 2} \! \hat X_\mathrm{sqz}-2)/4$ and
$\langle \hat n \rangle_\mathrm{thermal}=\langle \hat n
\rangle_\mathrm{total}-\langle \hat n \rangle_\mathrm{non-cl}$,
respectively, and the total mean photon number per bandwidth per
time is $\langle \hat n \rangle_\mathrm{total}=(\Delta^{\! 2} \!
\hat X+\Delta^{\! 2} \! Y-2)/4$ \cite{BOW03b}.
The effect of purification is clearly seen. The number of thermal
photons per bandwidth per time drops rapidly with only a very
small corresponding change in the number of non-classical photons.
One benefit of the photon number diagram is that efficacy contours
can be plotted for quantum information processes. In our case this
enables the tap-off fraction $\sqrt{1-\eta}$ to be chosen to ensure that
the quantum resources are optimally configured to suit the quantum
information protocol they will be utilized for.

The purification scheme investigated in this paper is particularly
useful for protocols where the total number of photons is a
critical resource.  One example of such is dense coding, where
entanglement is utilized to enhance the capacity of an information
channel to beyond the Holevo limit \cite{RAL02-DRU94}
$C_{\rm Holevo} \! \! = \! \left [ 1 \! + \! \langle \hat n
\rangle_{\rm cp} \right ] \log_{2} \left [ 1 \! + \! \langle \hat
n \rangle_{\rm cp} \right ]-\langle \hat n \rangle_{\rm cp} \log_{2}\langle \hat n \rangle_{\rm cp} $
where $\langle \hat n \rangle_{\rm cp}$ is the mean total number
of photons per bandwidth per time used in the communication
process. The channel capacity of the continuous variable dense coding protocol  \cite{braunstein-li} is given by \cite{BOW03b}
\begin{equation}
    C_{\rm EPR} \! = \log_{2} \! \! \left [ 1 \! + \frac{\langle \hat n \rangle_{\rm signal}}
    {2 \langle \hat n \rangle_{\rm non-cl} \! \! + \! 1 \! - \! 2 \sqrt{ \! \langle \hat n \rangle_{\rm
    non-cl}^2 \! \! + \! \langle \hat n \rangle_{\rm non-cl} }}
    \right ]
    \nonumber
\end{equation}
with $\langle \hat n \rangle_{\rm signal} = \langle \hat n \rangle_{\rm cp} - \langle \hat n \rangle_{\rm thermal} - \langle \hat n \rangle_{\rm non-cl}$ \footnote{We assume that the entanglement is produced by two identical squeezed beams. Therefore, the terms used here are related to those used in \cite{BOW03b} as follows: $2 \langle \hat n \rangle_{\rm non-cl} = \bar n_{\rm min}$, $2 \langle \hat n \rangle_{\rm thermal} = \bar n_{\rm excess}$, $\langle \hat n \rangle_{\rm cp} = \bar n_{\rm encoding}$, $2\langle n \rangle_{\rm total}=\bar{n}_{\rm total}$. The factors of two arise because photon numbers here are calculated for a single squeezed beam, whereas in \cite{BOW03b} they represent the entire entangled system.}. 
In the limit that $\langle \hat n \rangle_{\rm cp}  \! = \! \langle \hat n
\rangle_{\rm signal} \! + \!  \langle \hat n \rangle_{\rm total}
\! \gtrsim \! \langle \hat n \rangle_{\rm thermal}$, $C_{\rm EPR}$
is highly sensitive to the level of impurity ($\langle \hat n
\rangle_{\rm thermal}$) present in the system. Dense coding
channel capacity contours for $\langle \hat n \rangle_{\rm cp} =
400$ are shown in Fig.~\ref{PhotonNumberDiagram}.  The degradation
in channel capacity caused by thermal photons can be seen directly
from the slant of the contours.

Examining the dense coding efficacy contours near our experimental
data points, we see that with no purification the optimal channel
capacity achievable with our squeezing is $C_{\rm EPR} \approx
0.925 C_{\rm clas}$.  Employing the purification scheme this is
enhanced to $C_{\rm EPR} \approx 0.95 C_{\rm clas}$. Even though
the channel capacity is improved, the dense coding protocol
remains unsuccessful ($C_{\rm EPR} < C_{\rm clas}$).
However, the effectiveness
of our scheme when applied to dense coding can be judged from the
trajectories labelled b) in Fig.~\ref{PhotonNumberDiagram}.  In
this case, we consider squeezed resources with $\Delta^{\! 2} \!
\hat X_{\rm sqz} = 0.3$ and $\Delta^{\! 2} \! \hat  Y_{\rm
anti}=400$. The threshold for successful dense coding can then be
easily surpassed. Therefore, our purification scheme can be used
to transform quantum resources from a form for which dense coding
is not possible, into a form which allows for its successful
implementation.

Let us look at another interesting aspect of the purification
scheme. The entanglement generated by the interference of two
squeezed beams on an asymmetric beam splitter can be characterized
in terms of the entanglement of formation if the covariance matrix is transformed into the normal form~\cite{GIE03}. If a perfectly symmetric beam splitter is used to generate
entanglement, the entanglement of formation is independent of the
purity of the squeezed states. In realistic systems however, the beam splitting ratio is not perfectly symmetric and in that case, the generated correlations depend on the purity of the squeezed input states. For very asymmetric beam splitting ratios and for squeezed beams with
high excess noise levels we find that the degree of entanglement quite surprisingly can be
increased by purifying the squeezed resources despite the fact that less squeezing is actually available after purification. 

In summary, we have presented and demonstrated an optimal linear
optics and homodyne detection based purification  scheme to remove
thermal noise from squeezed states. Over an order of magnitude of
thermal noise reduction was achieved with a corresponding reduction
in squeezing of only 11~\%. This protocol is useful for quantum
information protocols such as dense coding and entanglement
swapping, and can enhance the levels of entanglement available in
realistic systems.

We thank T.~C.~Ralph and C.~Silberhorn for useful discussions. The work was supported by the EU project COVAQIAL (No. FP6-511004). R.F. acknowledges support from the projects GACR 202/03/D239, MSM6198959213 and the Alexander von Humboldt foundation. W.P.B. acknowledges funding from the MacDiarmid Institute.

\end{document}